\documentclass[11pt]{article}
\usepackage{amsmath,amssymb,amsfonts}
\usepackage{graphicx}
\usepackage{cite}
\usepackage{geometry}
\usepackage{comment}
\usepackage{authblk}
\usepackage{tikz}
\usepackage{mathrsfs}
\usepackage[outline]{contour}
\usetikzlibrary{decorations.markings,decorations.pathmorphing}
\usetikzlibrary{arrows.meta}
\contourlength{1.4pt}

\tikzset{>=latex}

\colorlet{myred}{red!80!black}
\colorlet{myblue}{blue!80!black}
\colorlet{mygreen}{green!70!black}
\colorlet{mydarkred}{red!50!black}
\colorlet{mydarkblue}{blue!50!black}
\colorlet{mypurple}{purple!70!black}
\colorlet{myorange}{orange!80!black}

\tikzset{
  arrow/.style={thick,->,>=stealth},
  dasharrow/.style={thick,dashed,->,>=stealth},
  lightcone/.style={thick,gray!60},
  horizon/.style={very thick},
  matter/.style={thick,mygreen},
  apparent/.style={thick,dashed,myred},
  event/.style={very thick,mydarkblue}
}

\title{
Adiabatic Anisotropic Gravitational Collapse in Painlevé-Gullstrand Coordinates: A Geometric Analysis
}

\author[1,2]{G. Abellán\thanks{gabriel.abellan@ciens.ucv.ve, gabriel@astrumdrive.com}}
\author[1,2]{N. Bolívar\thanks{nelson.e.bolivar@ucv.ve, nelson@astrumdrive.com}}
\author[2,3]{A. Alexandrova
}
\author[2]{I. Vasilev\thanks{ivaylo@astrumdrive.com}}

\affil[1]{Departamento de Física, Facultad de Ciencias, Universidad Central de Venezuela, 
Av. Los Ilustres, Caracas, 1041-A, Venezuela.}

\affil[2]{Astrum Drive Technologies, Dallas Pkwy Unit 120 B, Frisco, TX 75034, USA.}

\affil[3]{American College of Sofia, Floyd Black Lane, Mladost 2. 1799 Sofia, Bulgaria.}

\date{Received: date / Accepted: date}

\begin{document}

\maketitle

\begin{abstract}
We present a detailed geometric analysis of adiabatic, anisotropic gravitational collapse formulated in a single Painlevé–Gullstrand coordinate system that covers both the interior and exterior, thereby eliminating cross–chart matching artifacts. Building on the Oppenheimer–Snyder framework with a phenomenologically motivated energy–density profile, we enforce the Israel junction conditions and obtain closed-form surface evolution. Within this unified chart we derive exact solutions for the complete collapse process, characterize the causal structure, and track horizon formation and evolution. In particular, we identify and analyse a double apparent-horizon phase inside the matter and show that the event horizon stabilizes at the Schwarzschild radius. We further obtain critical parameter relations that govern the dynamics, including a threshold linking initial compactness to immediate horizon formation. The model is geometrically self-consistent within Einstein’s equations but exhibits violations of the standard point-wise energy conditions, highlighting known limitations of idealized anisotropic matter models and delineating the boundary where classical descriptions become inadequate. Together, these results provide geometric insights, compact analytic benchmarks and a didactic, coordinate-uniform perspective on collapse and horizon dynamics.
\end{abstract}

\section{Introduction}
\label{sec:intro}
Gravitational collapse represents one of the most fundamental processes in general relativity, governing the formation of black holes and determining the ultimate fate of massive astrophysical objects. The classical Oppenheimer--Snyder (OS) model \cite{Oppenheimer:1939ue} established the theoretical foundation for understanding spherically symmetric dust collapse, demonstrating that sufficiently massive objects inevitably form black holes when pressure forces cannot halt gravitational contraction. This seminal work revealed the essential physics of horizon formation and the inevitability of singularities in general relativity. However, realistic astrophysical collapse scenarios involve significant complications beyond the idealized dust model. Dense matter in neutron stars and stellar cores exhibits anisotropic pressure effects due to magnetic fields, phase transitions, and relativistic particle interactions \cite{Herrera1979a,Ruderman:1972aj,Usov:2004iz}. These anisotropies fundamentally alter the collapse dynamics and horizon formation process, potentially leading to qualitatively different outcomes compared to the uniform density case.

The mathematical treatment of anisotropic collapse has traditionally faced technical challenges related to coordinate singularities and the complexity of matching interior matter distributions to exterior vacuum solutions. Standard approaches often employ different coordinate systems for interior and exterior regions—typically comoving coordinates for the interior and Schwarzschild coordinates for the exterior—leading to computational difficulties and potential coordinate artifacts at the boundary \cite{Misner:1964je,Misner:1973prb,Chan:2014zra,Ivanov:2015dch,Chakrabarti:2024nyx}. These coordinate mismatches can obscure the physical interpretation of results and complicate the analysis of horizon formation across the stellar surface. 
Recent advances in numerical relativity have revealed rich geometric structures during gravitational collapse, including the formation of multiple apparent horizons and complex causal relations \cite{Hayward:1993mw,Ashtekar:2004cn,Nielsen:2005af,Gourgoulhon:2007ue,Alcubierre2008a,Baumgarte:2010ndz,Ivanov:2015mcw,Buoninfante:2024oyi}. 
The discovery of intricate horizon dynamics in numerical simulations suggests that simplified analytical models may miss essential features of the collapse process.

This work presents a specific analytical solution for anisotropic gravitational collapse using Painlevé-Gullstrand (PG) coordinates throughout both interior and exterior regions. 
The choice of PG coordinates is motivated by several compelling advantages. Unlike traditional approaches that couple different coordinate systems, PG coordinates provide a unified description of both interior matter and exterior vacuum regions. This eliminates coordinate singularities at the event horizon and avoids the technical complications associated with matching different metric forms across the stellar boundary \cite{Lasky:2006zz,Lasky:2007ky,Leonard:2011ce,Faraoni:2020ehi,Chakrabarti:2024nyx}. The ``river'' interpretation of PG coordinates, where the metric coefficient $\beta(t,r)$ represents the flow velocity of space itself, offers an intuitive physical picture that aligns naturally with the dynamics of gravitational collapse. 
Our unified coordinate approach enables us to investigate collapse scenarios that would be difficult to analyse using conventional methods. By maintaining the same coordinate structure throughout spacetime, we can track the formation and evolution of horizons without coordinate-dependent artifacts. 

To close the system of Einstein equations, we propose a phenomenologically motivated energy density profile that generalizes the Oppenheimer-Snyder model while preserving essential features of gravitational dynamics. The ansatz incorporates the $(t_c - t)^{-2}$ temporal scaling characteristic of homogeneous collapse models while introducing spatial inhomogeneity through a parabolic radial profile. 
The resulting stress-energy tensor exhibits anisotropic pressure components that emerge naturally from Einstein's equations rather than being imposed through phenomenological assumptions. This self-consistent approach ensures that the geometric and matter aspects of the model remain mutually compatible, even though the effective matter description may not correspond directly to specific astrophysical materials \cite{Noureen:2015nja,Pretel:2020xuo}. This matter content should be interpreted as an effective anisotropic source and the solution is intended as a toy model analytic benchmark for horizon dynamics.

The paper is structured to provide a complete analytical framework for anisotropic gravitational collapse, progressing from mathematical foundations through detailed results to physical interpretation. 
Section \ref{sec:theory} establishes our theoretical framework, beginning with the general Painlevé-Gullstrand line element and tetrad formalism that enables unified treatment of interior and exterior regions. We derive the Einstein tensor components using Cartan's method and establish the form of the anisotropic stress-energy tensor. The section concludes with an analysis of junction conditions and surface dynamics. We apply Israel's matching conditions to ensure smooth geometric connection between interior and exterior spacetimes, deriving the equations governing stellar surface motion during collapse. 
Section \ref{sec:horizons} develops the theory of null geodesics and horizon formation in our collapse geometry. We derive the conditions for apparent horizon formation and analyze the evolution of both interior and exterior horizons. This section reveals the double horizon phenomenon and establishes the critical parameter relationships that govern horizon dynamics. The event horizon analysis demonstrates how the global causal structure emerges from local geometric properties. 
Section \ref{sec:energy} provides a comprehensive analysis of energy conditions and their violations. We compute all stress-energy tensor components and systematically examine weak, null, dominant, and strong energy conditions. The discussion contextualizes these violations within the broader framework of gravitational collapse theory and explores their implications for the model's physical validity. 
Section \ref{sec:conclusions} presents our conclusions and discusses future research directions. We summarize the key contributions of our work and identify specific areas where extensions of the theoretical framework could provide additional insights into gravitational collapse phenomena.

\section{Theoretical Framework}
\label{sec:theory}

In this section, we establish the mathematical foundation for our gravitational collapse model. We employ Painlevé-Gullstrand coordinates and tetrad formalism to describe the dynamics of a spherically symmetric fluid collapsing under its own gravity. Unlike the classical treatment that uses different coordinate systems for interior and exterior regions, we develop a unified framework that smoothly describes both regions while avoiding coordinate singularities at the event horizon. We work using a flat local frame defined by an orthonormal tetrad, which allows us to apply the Cartan formalism for computing Einstein's field equations directly in flat frame. This provides analytical solutions for the complete gravitational collapse process, including the formation and evolution of apparent and event horizons.

\subsection{Spacetime Geometry}\label{subsec:geometry}

We consider a spherically symmetric spacetime with the general Painlevé-Gullstrand line element
\begin{equation}\label{eq:lineGeneral}
ds^2 = -dt^2 + (dr - \beta(t,r)dt)^2 + r^2 d\Omega^2\;,
\end{equation}
where $d\Omega^2 = d\theta^2 + \sin^2\theta d\phi^2$ is the metric on the unit 2-sphere, and $\beta(t,r)$ is a functions to be determined by Einstein's field equations. This metric form, introduced by Painlev\'{e} and Gullstrand \cite{Kanai2011}, represents the spacetime as a ``river'' where $\beta(t, r)$ encodes the inward flow velocity, providing 
an intuitive free-falling observer perspective that aligns with the dynamics of gravitational collapse 
in general relativity. A key advantage of employing Painlev\'{e} and Gullstrand coordinates throughout both interior and exterior regions is the elimination of coordinate mismatches that can arise when coupling different metric forms (e.g., comoving coordinates interior with Schwarzschild exterior). This unified approach ensures that apparent singularities due to coordinate choices are avoided, allowing for a cleaner analysis of the physical collapse dynamics across the matter boundary.

To avoid coordinate singularities and work in a physically meaningful frame, we introduce an orthonormal tetrad characterized by 1-forms $\{\omega^{\hat{a}}\}$ and vector fields $\{e_{\hat{a}}\}$ which provide a local flat frame 
\begin{align}
& \omega^{\hat{0}} = dt\;, \quad\quad \omega^{\hat{1}} = dr - \beta(t,r)dt\;, \quad\quad 
\omega^{\hat{2}} = rd\theta\;, \quad\quad \omega^{\hat{3}} = r\sin{\theta}d\varphi\;, \\
& \partial_{\hat{0}} = \partial_t + \beta(t,r)\partial_r \;,  \quad\quad
\partial_{\hat{1}} = \partial_r \;, \quad\quad
\partial_{\hat{2}} = \frac{1}{r} \partial_\theta \;, \quad\quad
\partial_{\hat{3}} = \frac{1}{r\sin\theta} \partial_\varphi\;.
\end{align}
Here, $\beta(t,r)$ encodes the dynamics of the collapsing matter and represents the velocity of freely falling observers comovil with the collapsing fluid.

The non-vanishing components of the Einstein tensor are computed using the structure equations
\begin{align}
d\omega^{\hat{a}} + \omega^{\hat{a}}{}_{\hat{b}} \wedge \omega^{\hat{b}} &= 0, \\
d\omega^{\hat{a}}{}_{\hat{b}} + \omega^{\hat{a}}{}_{\hat{c}} \wedge \omega^{\hat{c}}{}_{\hat{b}} &= \Omega^{\hat{a}}{}_{\hat{b}},
\end{align}
where $\omega^{\hat{a}}$ are the basis 1-forms, $\omega^{\hat{a}}{}_{\hat{b}}$ are the connection 1-forms, and $\Omega^{\hat{a}}{}_{\hat{b}}$ are the curvature 2-forms.

Using the Cartan formalism, the Einstein tensor $G_{\hat{a}\hat{b}}= R_{\hat{a}\hat{b}}-\frac{1}{2}\eta_{\hat{a}\hat{b}}R$ in the local flat frame take the following form
\begin{eqnarray}
    G_{\hat{t}\hat{t}} &=& 
    \frac{\beta}{r^2}\Bigl(\beta + 2r\,\beta'\Bigr)\;, \label{eq:EinTen01}  \\
    G_{\hat{r}\hat{r}} &=& -\,\frac{\beta}{r^2}\Bigl(\beta + 2r\,\beta'\Bigr)
        + 2\frac{\dot{\beta}}{r} \;, \label{eq:EinTen02} \\
    G_{\hat{\theta}\hat{\theta}} &=& 
	-\,\beta\,\beta'' \;-\; (\beta')^2 \;-\; 2\,\frac{\beta}{r}\,\beta' 
        + \frac{\dot{\beta}}{r} + \dot{\beta}' \;\;=\;\; G_{\hat{\varphi}\hat{\varphi}} \;. \label{eq:EinTen03}
	\end{eqnarray}
Here the dot represents time derivatives and prime radial derivatives. As expected, spherical symmetry imposes $G_{\hat{\theta}\hat{\theta}}=G_{\hat{\varphi}\hat{\varphi}}$. All other components of $G_{\hat{a}\hat{b}}$ vanish.

\subsubsection{Matter Description}
Given the form of the Einstein tensor, the most general matter content we can include is an anisotropic fluid with stress-energy tensor
\begin{equation}\label{eq:interiorT}
T_{\hat{a}\hat{b}} = \text{diag}(\rho, p_r, p_\perp, p_\perp),
\end{equation}
where $\rho(t,r)$ is the energy density, $p_r(t,r)$ is the radial pressure, and $p_\perp(t,r)$ is the tangential pressure. The anisotropy arises naturally from the geometry. Therefore we will model the adiabatic collapse of an anisotropic fluid with spherical symmetry. Anisotropic fluids are physically motivated in high-density collapse scenarios, such as neutron stars with phase transitions or magnetic fields, where radial and tangential pressures differ \cite{Herrera1979a}. In adiabatic models, this anisotropy arises naturally from relativistic effects without dissipation.

For the external spacetime, on the other hand, we will consider space to be vacuum, free of both matter and radiation $T_{\hat{a}\hat{b}}=0$. From this point on we remove the hat on the indices because we will always be working in the local flat frame.

\subsection{Interior and Exterior Solutions}\label{subsec:solutions}
Now, we will move on to find and solve the Einstein equations in each region. To do so, we will be guided by a discussion based on the phenomenology of the problem and the interpretation of the line element \eqref{eq:lineGeneral}.

\subsubsection{Interior Solution ($r<R(t)$)}

The interior of the collapsing distribution is modeled as an adiabatic anisotropic medium described by Painlevé-Gullstrand coordinates, this takes the form
\begin{equation}
ds^2_{<} = -dt^2 + (dr - \beta_{<}(t,r) dt)^2 + r^2 d\Omega^2 \;,
\end{equation}
where $\beta_{<}(t,r)$ describes the velocity field along the entire fluid. This function determines the interior spacetime curvature. Using the energy-momentum tensor \eqref{eq:interiorT} we see that Einstein's equations for the inner region are given by
\begin{align}
8\pi \rho &= \frac{1}{r^2} 
\frac{\partial}{\partial r}\big(r\beta_{<}^2\big)\;, \label{eq:EeqInt01} \\
8\pi p_r &= -\frac{1}{r^2} \frac{\partial}{\partial r}\big(r\beta_{<}^2\big)
+ 2\frac{\dot{\beta}_{<}}{r} \;, \label{eq:EeqInt02} \\
8\pi p_\perp &= -\,\beta_{<}\,\beta''_{<} \;-\; (\beta'_{<})^2 \;-\; \frac{2}{r}\,\beta_{<}\beta'_{<} 
        + \frac{\dot{\beta}_{<}}{r} + \dot{\beta}'_{<} \;. \label{eq:EeqInt03}
\end{align}
Fluid systems usually obey equations of state. For this system this relationship is not evident.

The above system of equations has four functions, therefore an additional condition is needed in order to close the system and find a solution. 
In this work we propose an energy density inspired by the Oppenheimer--Snyder model of dust collapse
\begin{equation} \label{eq:densityInt}
\rho(t,r) = \frac{\alpha_c}{(t_c-t)^{2}}\left[1-\left(\frac{r}{R(t)}\right)^{\!2}\right]\;,
\end{equation}
where $\alpha_c$ is a normalization constant that depends on the total mass of the system, and $R(t)$ is the surface radius. In addition, notice that the collapse starts at $t = 0$ and ends at the collapse time $t_c$. This form generalizes the uniform dust density in the Oppenheimer--Snyder
model \cite{Oppenheimer:1939ue} to a non-uniform profile, allowing for anisotropic pressures while preserving the $(t_c - t)^{-2}$ time dependence observed in classical collapse \cite{Kanai2011}.

The choice of this specific functional form requires justification beyond mathematical convenience. The $(t_c - t)^{-2}$ scaling is well-established in homogeneous collapse models and represents the natural time evolution for gravitational systems approaching a singularity. The spatial profile $[1 - (r/R(t))^2]$ ensures finite density at the center with smooth variation to the boundary, which is physically reasonable for a self-gravitating fluid. While not derived from a specific microphysical equation of state, this ansatz captures essential features of gravitational collapse: finite central density, smooth spatial variation, and the correct temporal scaling. The resulting anisotropic pressures emerge naturally from Einstein's equations rather than being imposed \emph{a priori}, making this a self-consistent gravitational model despite its phenomenological nature. This source may represent an effective coarse-grained model of a fundamental microscopic model, but those details are beyond the scope of the present work.

Taking the expression \eqref{eq:densityInt} for the energy density we can integrate equation \eqref{eq:EeqInt01} and get
\begin{align}
\beta^2_{<}(t,r) &= \frac{r_0\beta^2(t,r_0)}{r} + \frac{8\pi}{r}\!\int_{r_0}^r \tilde{r}^2 \rho(t,\tilde{r})\,d\tilde{r} \nonumber \\
&= \frac{8\pi\alpha_c R^2(t)}{3(t_c-t)^{2}}
\Big( \frac{r}{R(t)} \Big)^{\!2}
\left[1-\frac{3}{5}\left(\frac{r}{R(t)}\right)^{\!2}\right]\,. \label{eq:IntSol}
\end{align}
The integration is performed by considering $r$ inside the distribution surface. It is important to mention that in order to consider the collapse we choose the negative sign when taking the square root.

\subsubsection{Exterior Solution ($r\geq R(t)$)}

The exterior geometry is described by a static, spherically symmetric metric in Painlevé-Gullstrand form:
\begin{equation}
ds^2_{>} = -dt^2 + \left(dr - \beta_{>}(r) dt\right)^2 + r^2d\Omega^2,
\end{equation}
By also working in PG coordinates for the exterior, it is possible to avoid the occurrence of singularities that come from coupling different types of metrics for the interior and exterior regions. We expect a smooth description of both the exterior vacuum region and the transition across the horizon.

For simplicity, we assume vacuum in the exterior spacetime.  This is a good approximation for adiabatic collapse models. Taking this into account, we write the equations for the outer region
\begin{align}
0 &= \frac{1}{r^2} 
\frac{\partial}{\partial r}\big(r\beta_{>}^2\big)\;, \label{eq:EeqExt01} \\
0 &= \beta_{>}\,\beta''_{>} \;+\; (\beta'_{>})^2 \;+\; \frac{2}{r}\,\beta_{>}\beta'_{>} 
        \;. \label{eq:EeqExt02}
\end{align}
The vacuum assumption is a standard approximation for adiabatic collapse \cite{Oppenheimer:1939ue}, though real systems may include radiation, potentially altering horizon formation times as noted in dissipative models \cite{Penrose:1969pc}.

As expected, since we are looking for a static exterior solution, the answer for this system corresponds to the Schwarzschild solution expressed in Painlevé--Gullstrand coordinates
\begin{align}
\beta^2_{>}(r) &= \frac{2M_0}{r} \,. \label{eq:ExtSol}
\end{align}
This leads to a model where the Schwarzschild mass $M_0$ associated with the distribution remains fixed during the whole process. When taking the square root we choose the negative sign in order to describe collapse in a consistent way. 

\begin{figure*}[htbp]
\centering
\begin{tabular}{cc}
\hspace{-.5cm}
\includegraphics[width=0.41\textwidth]{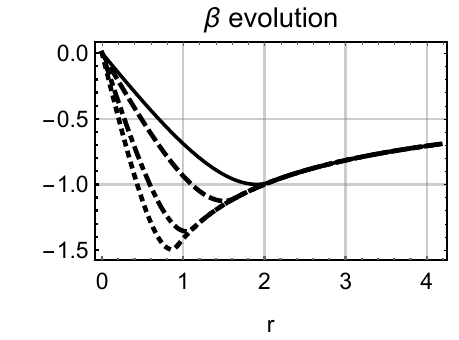} & \hspace{.5cm}
\includegraphics[width=0.41\textwidth]{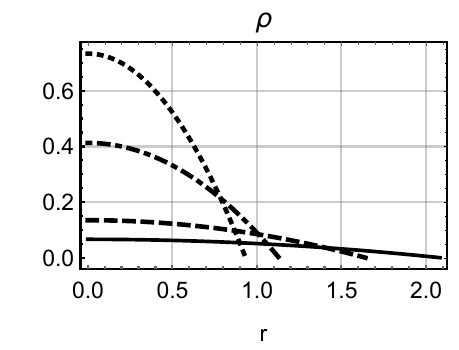} 
\end{tabular}
\caption{\textbf{(left)} Evolution of the metric function $\beta$ vs. $r$ for several values of time $t$. The difference inside matter (growing towards negative values) and outside matter (decaying towards zero) is clearly seen. \textbf{(right)} Energy density for various values of $t$. Parameters used, $M_0=1$. The selected values are $t=0$ (solid line), $t=0.3 t_c$ (dashed line), $t=0.6t_c$ (dot--dashed line), $t=0.7t_c$ (short--dashed line).}
\label{fig:Beta_Densidy}
\end{figure*}

\subsection{Israel Junction Conditions and Surface Dynamics}\label{subsec:surface}

The dynamics of the stellar surface are governed by the requirement of smooth matching between interior and exterior geometries. At the surface, we impose Israel junction conditions \cite{Israel1966a}, which demand continuity of both the first and second fundamental forms.

\subsubsection{First Fundamental Form and Surface Radius $R$}
The first fundamental form for the induced metric $h_{ab}$ is obtained by evaluating the spacetime metric on the tangent vectors to the hypersurface $\Sigma$. The non-trivial temporal component yields
\begin{equation}
h_{tt}^\pm = -1 +(\beta^{\pm} - \dot{R})^2 \;,
\end{equation}
where $\beta^- = \beta_{<}(t, R(t))$ and $\beta^+ = \beta_{>}(R(t))$. The angular components are identical on both sides:
\begin{equation}
h_{\theta\theta}^\pm = R(t)^2, \quad h_{\phi\phi}^\pm = R(t)^2 \sin^2\theta\;.
\end{equation}
The Israel junction condition for the first fundamental form requires $h_{ab}^- = h_{ab}^+$, which leads to the following non--trivial continuity condition
\begin{equation}
\beta_{<}(t, R(t)) = \beta_{>}(R(t))\;.
\end{equation}
This condition ensures that the shift function is continuous across the collapse surface, preventing unphysical discontinuities in the coordinate flow. This expression provides a relation that allows to obtain the resulting dynamics for the surface of the fluid. Using the expressions \eqref{eq:IntSol} and \eqref{eq:ExtSol}, we find
\begin{equation}\label{eq:radius}
    \frac{2M_0}{R(t)}=\frac{4R(t)^2}{9(t_c-t)} \quad\longrightarrow\quad
    R(t)^3 = \frac{9}{2}M_0(t_c-t)^2\;.
\end{equation}
Note that the radius $R(t)$ vanishes in finite time $t_c$. This is expected from ansatz \eqref{eq:densityInt} for the energy density. In order to ensure consistency with previous results \eqref{eq:densityInt} and \eqref{eq:IntSol} for an adiabatic collapse model, we require the total mass of the distribution to be constant throughout the process
\begin{equation}
\int_0^{R(t)} 4\pi r^2 \frac{\alpha_c}{(t_c-t)^{2}}   
\left[1-\left(\frac{r}{R(t)}\right)^{\!2}\right] dr = M_0\;.
\end{equation}
Substituting $R(t)$ in this expression, leads to $\alpha_c=5/12\pi$. This value will be used from now on.

To conclude this section, it is important to note that there is a deep connection between metric function $\beta(t,r)$ and $R(t)$. To clearly establish this relation let us take the expression \eqref{eq:radius} for $R(t)$ and calculate its time derivative
\begin{equation}
    3R^2(t)\dot{R}(t) = -9M_0(t_c-t) 
    \quad \longrightarrow \quad 
    \dot{R}^2(t) = \frac{9M^2_0}{R^4(t)}(t_c-t)^2 \,.
\end{equation}
Using the equation for $R(t)$ and simplifying, we arrive at the following result
\begin{equation}
    \dot{R}^2(t) = \frac{2M_0}{R(t)} = \beta^{2}_{>}(t,R(t)) 
    = \beta^{2}_{<}(t,R(t)) \;.
\end{equation}
Here we have used the already established continuity of $\beta(t,r)$ at $r=R(t)$. What we just find is that metric function $\beta(t,r)$ is related to the collapse velocity $\dot{R}(t)$ and therefore at $r=R(t)$ it coincides with the velocity of the collapsing surface. This is reasonable since the Painlev\'e--Gullstrand coordinates are adapted to an observer freely falling with matter so that this result serves as a cross check of the consistency of the model. The immediate consequence is 
\begin{eqnarray}
    ds^2_{\Sigma} = -dt^2 + R^2(t)d\theta^2 + R^2(t)\sin^2\theta d\varphi^2 \;,
\end{eqnarray}
that is, the line element induced on the hypersurface $r=R(t)$ is flat.

All these equations encode the fundamental dynamics of gravitational collapse and determine the evolution of the surface from its initial configuration to the possible formation of the event horizon.

\subsubsection{Second Fundamental Form}

To complete the Israel junction conditions, we examine the second fundamental form $K_{ab}$ at the hypersurface $\Sigma:$ $r = R(t)$. 
In a local flat frame, the metric takes the Minkowski form $\eta_{ab} = \text{diag}(-1, 1, 1, 1)$, and all geometric calculations are performed using this flat metric. 
The hypersurface $\Sigma$ 
represents surfaces of instantaneous constant radial coordinate. In the local flat frame, the normal vector to such a surface is simply
\begin{equation}
n^a = (0, 1, 0, 0)\;,
\end{equation}
this represents the direction orthogonal to the spatial hypersurface in the instantaneous rest frame. In the tetrad basis, $e_1 = \partial_r$ points in the radial direction 
which is the direction of increasing $r$. 
From this expression, the 1-forms and dual basis adapted to the hypersurface $\Sigma$ is given by
\begin{eqnarray}
    &\theta^{0} = dt\,, \quad 
    \theta^2 = R(t)d\theta\,, \quad 
    \theta^3 = R(t)\sin{\theta}d\varphi\,, \label{eq:surface_tetrad}\\
    & e_{0} = \partial_t\,, \quad
    e_2 = \dfrac{1}{R(t)}\partial_\theta\,, \quad 
    e_3 = \dfrac{1}{R(t)\sin{\theta}}\partial_\varphi\,. \label{eq:surface_dual} 
\end{eqnarray}
With these adapted 1-forms, the Cartan spin connections with normal direction 1 are given by
\begin{align}
\Theta^1_{\;\;0} = - \dot{R}(t)\beta'(t,R(t))\, \theta^0 \;, \quad
\Theta^1_{\;\;2} = -\frac{1}{R(t)}\, \theta^2 \;, \quad
\Theta^1_{\;\;3} = -\frac{1}{R(t)}\, \theta^3\;.
\end{align}
From these expressions, the non-zero components of the extrinsic curvature are
\begin{align}
K_{00} &= \Theta^1_{\;\;0}(e_0) 
= -\dot{R}(t)\beta'(t,R(t)) \theta^0(e_0) 
= -\dot{R}(t)\beta'(t,R(t)) \;, \label{eq:K00}\\
K_{22} &= \Theta^1_{\;\;2}(e_2) 
= -\frac{1}{R(t)} \theta^2(e_2) 
= -\frac{1}{R(t)} \;,\label{eq:K22}\\
K_{33} &= \Theta^1_{\;\;3}(e_3) 
= -\frac{1}{R(t)} \theta^3(e_3)
= -\frac{1}{R(t)} \;. \label{eq:K33}
\end{align}
Note that, for the time direction, the extrinsic curvature value will be conditioned by the factor $\beta(t,r)\beta'(t,r)$ at $\Sigma$. Consider the general expression in \eqref{eq:IntSol} and calculating the $r$-derivative we find
\begin{eqnarray}
    \beta(t,r)\beta'(t,r) &=& -\frac{r_0 \beta^2(t,r_0)}{2r^2} -
     \frac{4\pi}{r^2}\,\int_{r_0}^r \tilde{r}^2 \rho(t,\tilde{r})\,d\tilde{r} +
     4\pi r \rho(t,r) \;,
\end{eqnarray}
and evaluating this expression on both sides of the hypersurface we get
\begin{eqnarray}
    \dot{R}(t)\beta'_{<}(t,R(t)) &=& 
    4\pi R(t) \rho_{<}(t,R(t)) - \frac{M_0}{R^2(t)} = 0 \;, \label{eq:der_disc01} \\
    \dot{R}(t)\beta'_{>}(t,R(t)) &=& 
    - \frac{M_0}{R^2(t)} + 4\pi R(t) \rho_{>}(t,R(t)) = 0 \,. \label{eq:der_disc02}
\end{eqnarray}
For $r<R(t)$ we set $r_0 = 0$ and the integral is related with total mass $M_0$, while for $r>R(t)$ we have $r_0 = R(t)$ and the integral vanish. Remarkably, both expressions cancel out because the energy density is null at the surface. This conclusion will be crucial for determining whether thin shells are present in the systems.

Israel formalism relates discontinuities in extrinsic curvature to the surface energy--momentum tensor
\begin{equation}\label{eq:jump_shell}
S_{ab} = -\frac{1}{8\pi}\left([K_{ab}] - \tilde{\eta}_{ab}[K]\right)
\end{equation}
where $[K_{ab}] = K_{ab}^{+} - K_{ab}^{-}$ denotes the jump across $r = R(t)$, $[K] = [K^{c}_{\;\;c}]$ is the trace of the jump, and $\tilde{\eta}_{ab}$ is the induced metric on the hypersurface. Based on expressions \eqref{eq:K00}--\eqref{eq:K33}, we find that the non-zero components are
\begin{eqnarray}
    & K_{00}-\tilde{\eta}_{00}K = 
    -\dfrac{2}{R(t)}\;, \label{eq:jump_00} \\
    & K_{22}-\tilde{\eta}_{22}K = 
    \dfrac{1}{R(t)} = K_{33}-\tilde{\eta}_{33}K\;, \label{eq:jump_22}
\end{eqnarray}
where we have used the equations \eqref{eq:der_disc01} and \eqref{eq:der_disc02}. It is worth stressing that we are working on a local flat base adapted to the hypersurface and therefore $\tilde{\eta}_{ab} = \text{diag(-1,1,1)}$. From this analysis we conclude that the model has no thin shells and is free of singularities in the matter sector. This can be seen in plot \ref{fig:Beta_Densidy} where it is realized that the metric function is continuous and smooth over the whole plotted domain.

This formulation provides the cleanest expression of the gravitational junction conditions for spherically symmetric collapse in coordinates adapted to the matter motion.

\section{Null Geodesics and Horizon Formation}
\label{sec:horizons}

In this section, we derive the null geodesic equations for Painlevé-Gullstrand coordinate system and analyse the resulting causal structure. The formation and evolution of horizons during gravitational collapse is analysed by examining the behaviour of null geodesics and applying the marginally trapped surface condition. We distinguish between apparent horizons, which are local geometric features, and the event horizon, which represents the global causal structure of the spacetime.

\subsection{Null Geodesics in the Collapse Geometry}

For the metrics under consideration, radial null geodesics satisfy the condition $ds^2 = 0$ with $d\theta = d\phi = 0$. This yields
\begin{equation}
-dt^2 + (dr - \beta dt)^2 = 0,
\end{equation}
which gives two classes of null directions
\begin{align}
\frac{dr}{dt} = 1 + \beta \quad \text{(outgoing)}\;, \quad\quad
\frac{dr}{dt} = -1 + \beta \quad \text{(ingoing)}\;. \label{eq:geodesic_null}
\end{align}
Remember that for collapse phenomena we have $\beta\leq 0$. The expansion scalar $\theta$ of the outgoing null congruence in spherically symmetric spacetimes is given by
\begin{equation}
\theta = \frac{2}{r} \frac{dr}{dt} = \frac{2}{r}(1 + \beta)\;. \label{eq:expansion_scalar}
\end{equation}
Applying conditions on the scalar expansion we find the regions where light can still escape.

\subsection{Apparent Horizon Condition}

An apparent horizon is a marginally outer trapped surface where the expansion of outgoing null geodesics vanishes $\theta = 0$. From equations \eqref{eq:geodesic_null} and \eqref{eq:expansion_scalar} we find
\begin{equation}
1 + \beta = 0 \quad\longrightarrow\quad \beta = -1\;. \label{eq:apparent_horizon_condition}
\end{equation}
In principle, this condition must be solved separately for the interior and exterior regions using the respective metric functions $\beta_{<}(t,r)$ and $\beta_{>}(r)$.

\subsubsection{Interior Apparent Horizon Evolution}

Using expression \eqref{eq:IntSol} for the interior solution ($r < R(t)$), the apparent horizon condition \eqref{eq:apparent_horizon_condition} becomes
\begin{equation}
r_{\text{AH}}^{\pm} = \sqrt{
\frac{5}{6}R(t)\bigg( 1 \pm \sqrt{1- \frac{12 R(t)}{25M_0}} \bigg)\;,
} \label{eq:interior_ah_condition}
\end{equation}
with $R(t)$ given by equation \eqref{eq:radius}. 

At this point several important observations merit discussion. First, equation \eqref{eq:apparent_horizon_condition} for apparent horizon formation yields two distinct solutions, denoted as $r_{AH}^{(+)}(t)$ and $r_{AH}^{(-)}(t)$. The existence of multiple apparent horizons during collapse is not unique to our model—similar phenomena have been reported in other anisotropic collapse studies. For instance, it has been found multiple marginally trapped surfaces in inhomogeneous dust collapse \cite{Goswami:2002ds}, while in \cite{Herrera1979a} was observed complex horizon structures in anisotropic perfect fluid models. This multiplicity reflects the rich geometric structure that can emerge in non-trivial collapse scenarios and should not be dismissed as purely mathematical artifact (see \cite{Joshi:2011rlc,Faraoni:2015ula,Cardoso:2019rvt,Terno:2020tsq}).

The physical validity of each solution requires verification that $r_{AH}^{(i)}(t) < R(t)$ during the collapse evolution. In figure 2, we observe how both apparent horizon solutions evolve relative to the matter surface radius. The outer solution $r_{AH}^{(+)}(t)$ grows with time until it reaches the surface, while the inner solution $r_{AH}^{(-)}(t)$ remains well within the matter distribution throughout the collapse.

Second, at the critical time $t_H$ when $r_{AH}^{(+)}(t_H) = R(t_H)$, the apparent horizon reaches the matter surface. From the junction conditions and the requirement that the exterior apparent horizon coincides with the Schwarzschild radius, we have
\begin{equation}
R(t_H) = r_{\text{AH}}^{(+)}(t_H) = 2M_0\;. \label{eq:critical_condition}
\end{equation}
For $t > t_H$, the solution $r_{\text{AH}}^{(+)}(t)$ becomes unphysical as it would correspond to $r > R(t)$ and is beyond the $\beta_{<}$ domain, leaving only the inner apparent horizon $r_{\text{AH}}^{(-)}(t)$ within the matter.

\begin{figure*}[htbp]
\centering
\begin{tabular}{cc}
\hspace{-.5cm}
\includegraphics[width=0.45\textwidth]{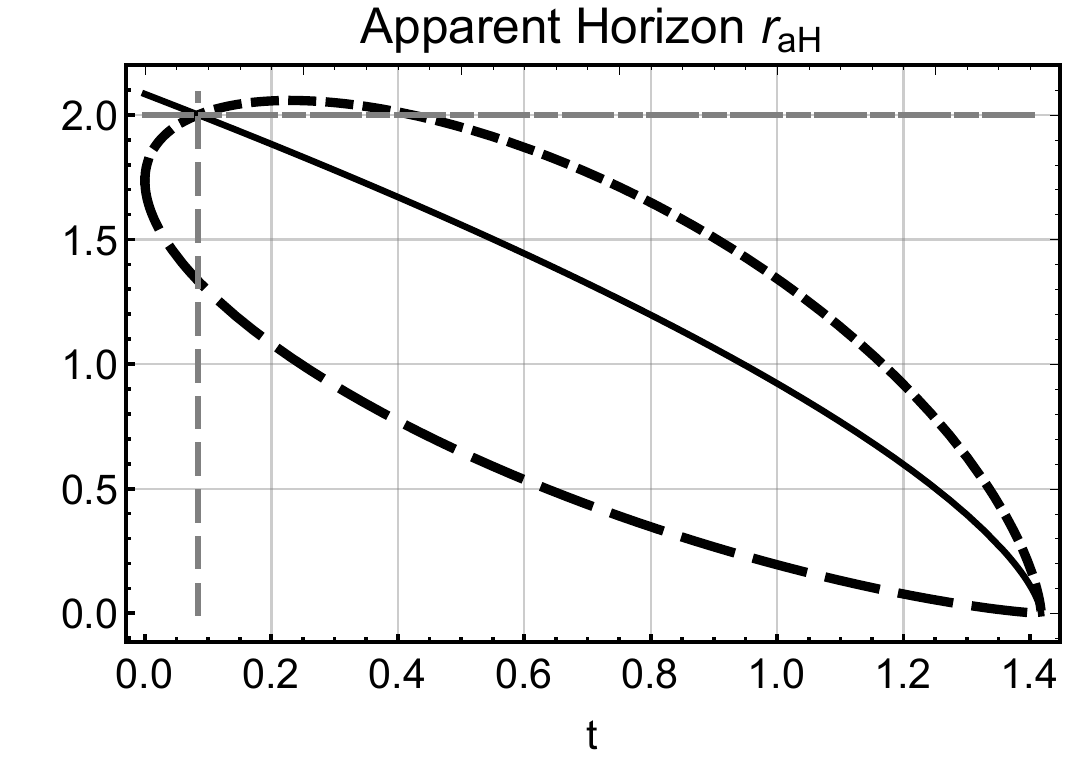} & \hspace{.5cm}
\includegraphics[width=0.45\textwidth]{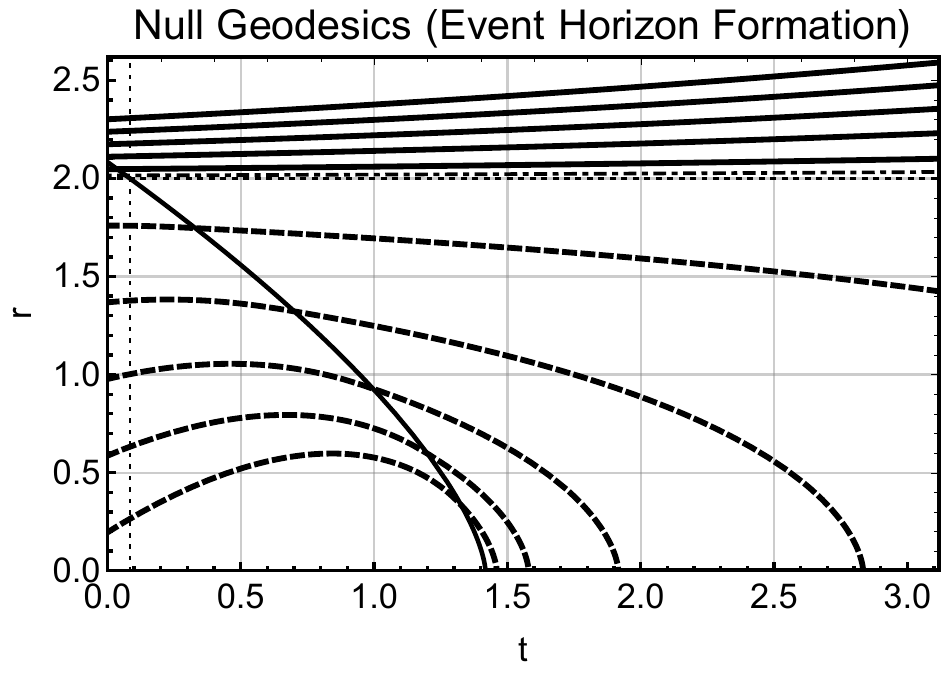} 
\end{tabular}
\caption{\textbf{(left)} The formation of two apparent horizons is observed at $t=0$. When the apparent horizon $r^{(+)}_{\text{AH}}$ (short dashed line) at $t=t_H$ reaches the surface of the distribution (solid line), it acquires the critical value $2M_0$ which coincides with the Schwarzschild radius. The apparent horizon $r^{(-)}_{\text{AH}}$ (long dashed line) is always within the distribution. \text{(right)} Several null geodesics are plotted. The dashed line indicates trapped geodesics. The solid line indicates escaping geodesics. A critical geodesic (horizontal dot--dashed line) is also observed right at the $2M_0$ value. The curve for the surface is included for comparison.}
\label{fig:horizon_evolution}
\end{figure*}

\subsubsection{Exterior Apparent Horizon}

In the exterior region ($r > R(t)$), which correspond to a Schwarzschild geometry with mass $M_0$, we have
\begin{equation}
\beta_{>}(r_{\text{AH}}^{\text{ext}}) = -1 
\quad \longrightarrow \quad 
r_{\text{AH}}^{\text{ext}} = 2M_0\;. \label{eq:exterior_ah}
\end{equation}
The external apparent horizon coincides with Schwarzschild's radius.

These horizon dynamics underscore the utility of PG coordinates in tracking null geodesics across the horizon without singularities, as demonstrated in numerical scalar field collapses, though energy violations in our model may exaggerate the role of anisotropy in preventing naked singularities.

\subsection{Event Horizon Analysis}

The event horizon is defined as the boundary of the causal past of future null infinity. For a collapse scenario leading to a Schwarzschild black hole with mass $M_0$, the event horizon is uniquely determined and coincides with the Schwarzschild radius:
\begin{equation}
r_{\text{EH}} = 2M_0. \label{eq:event_horizon}
\end{equation}
Unlike apparent horizons, which can exhibit complex temporal evolution and multiplicity, the event horizon remains at a fixed radial coordinate throughout the entire collapse process \cite{Thornburg:2006zb,Visser:2008cjw}.

\subsection{Horizon Dynamics Summary}

In general, in this model, the horizon structure during collapse exhibits the following phases:
\begin{enumerate}
\item \textbf{Early Phase} ($t < t_{\text{formation}}$): No apparent horizons exist within the matter.
\item \textbf{Double Horizon Phase} ($t_{\text{formation}} < t < t_H$): Two apparent horizons coexist within the matter: $r_{\text{AH}}^{(-)}(t) < r_{\text{AH}}^{(+)}(t) < R(t)$.
\item \textbf{Transition} ($t = t_H$): The outer apparent horizon reaches the surface: $r_{\text{AH}}^{(+)}(t_H) = R(t_H) = 2M_0$.
\item \textbf{Post-Transition Phase} ($t > t_H$): Only the inner apparent horizon $r_{\text{AH}}^{(-)}(t)$ remains within the matter, while the exterior spacetime maintains an apparent horizon at $r = 2M_0$.
\end{enumerate}
The transition from a double apparent horizon structure to a single horizon configuration represents a distinctive feature of the collapse dynamic presented in this work, highlighting the rich geometric structure that emerges during the formation of black holes in this model.

\subsection{Example}

For concrete analysis, we establish a specific parameter relationship that eliminates arbitrariness while highlighting key physical features. We choose the critical condition where apparent horizons first emerge, which occurs when the discriminant in equation \eqref{eq:interior_ah_condition} vanishes. The initial radius of the distribution is given by
\begin{equation}\label{eq:Initial-Radius}
    R_0^3 = \frac{9}{2}M_0t_c^2\;.
\end{equation}
This relation connects three fundamental parameters: the initial size $R_0$, the total mass $M_0$, and the collapse timescale $t_c$. To determine their specific values, we impose the physically motivated constraint that apparent horizons form at the onset of collapse ($t = 0$), which occurs when
\begin{equation}\label{eq:horizon-constraint}
    1-\frac{12R_0}{25M_0}=0 \quad\longrightarrow\quad
    R_0=\frac{25}{12}M_0\;.
\end{equation}
This critical condition is not arbitrary but represents a natural boundary case in the parameter space. For $R_0 > 25M_0/12$, apparent horizons would form at some time $t > 0$ during the evolution, while for $R_0 < 25M_0/12$, no real solutions exist for apparent horizons within the matter. The choice $R_0 = 25M_0/12$ therefore represents the threshold configuration where horizon formation is incipient.

Several physical insights emerge from this choice. First, $R_0 > 2M_0$, confirming that the Schwarzschild radius lies initially within the matter distribution which is a necessary condition for black hole formation through collapse. Second, this constraint provides a natural relationship between the initial compactness of the object and its total mass, suggesting that only sufficiently compact configurations can undergo horizon formation from the outset. With this constraint, the collapse time becomes fully determined
\begin{equation}\label{eq:collapse-time}
    t_c = \frac{125}{36\sqrt{6}}M_0\;,
\end{equation}
providing the complete parameter set needed for quantitative analysis. This choice enables systematic study of horizon dynamics while ensuring the model represents a physically meaningful limiting case rather than an arbitrary mathematical construction.

\section{Energy Conditions and Physical Validity}
\label{sec:energy}

Energy conditions constrain the properties of matter and energy in general relativity. These conditions encapsulate our intuitive expectations about the behavior of physical matter, such as positive energy density and causal relationships between energy, momentum, heat flow, etc. In this section, we derive the energy--momentum tensor components from Einstein's field equations and systematically examine whether the matter configuration in our generalized Oppenheimer-Snyder model satisfies the standard energy conditions.

\subsection{Energy--Momentum Tensor Components}
\label{subsec:stress_energy}

The matter content for the collapsing model is derived directly from Einstein's field equations. As we showed in section \ref{sec:theory}, the energy--momentum tensor takes the anisotropic perfect fluid form $T_{\hat{a}\hat{b}} = \text{diag}(\rho, p_r, p_\perp, p_\perp)$, 
with $\rho(t,r)$ is the energy density, $p_r(t,r)$ is the radial pressure, and $p_\perp(t,r)$ is the tangential pressure.

\subsubsection{Energy Density}
As we have already seen, the proposed energy density is an extension of the Oppenheimer--Snyder model. Using the value of $\alpha_c$ found previously, we have that
\begin{equation} \label{eq:densityIntDef}
\rho(t,r) = \frac{5}{12\pi(t_c-t)^{2}}\left[1-\left(\frac{r}{R(t)}\right)^2\right]\;.
\end{equation}
The density remains positive throughout the collapse, with radius $R(t)$ given by \eqref{eq:radius}.

\subsubsection{Pressure Components}
\begin{figure*}[htbp]
\centering
\begin{tabular}{ccc}
\hspace{-1.1cm}
\includegraphics[width=0.36\textwidth]{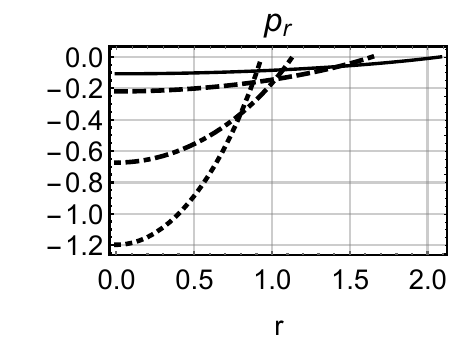} &
\hspace{-.6cm}
\includegraphics[width=0.36\textwidth]{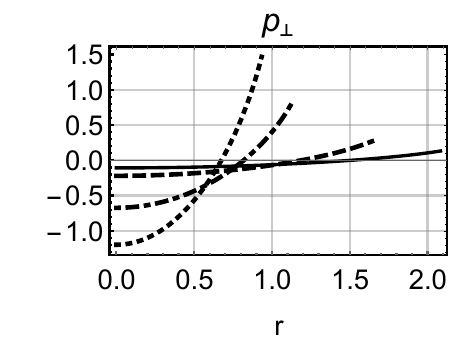} & \hspace{-.6cm}
\includegraphics[width=0.36\textwidth]{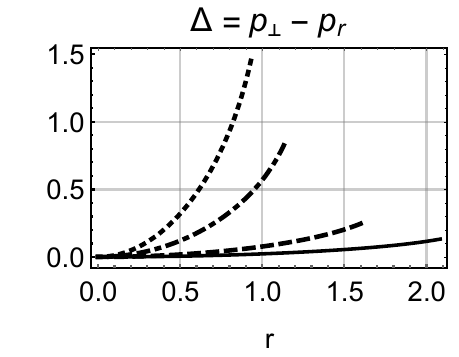}
\end{tabular}
\caption{\textbf{(left)} Radial Pressure. \textbf{(center)} Tangential pressure. \textbf{(right)} Anisotropy factor. Parameters used, $M_0=1$. The selected values are $t=0$ (solid line), $t=0.3 t_c$ (dashed line), $t=0.6t_c$ (dot--dashed line), $t=0.7t_c$ (short--dashed line).}
\label{fig:Solution_Matter}
\end{figure*}

An important aspect of the proposed model is that it provides us with the ability to model the collapse of an anisotropic fluid. We can calculate the radial and tangential components of the pressure using the Einstein equations. 
The radial pressure, derived from the $(\hat{1},\hat{1})$ component of Einstein's equations, is
\begin{equation}
p_r(t,r) = - \frac{5}{12\pi(t_c-t)^2}
\bigg[1 - \Big(\frac{r}{R(t)} \Big)^{\!2}\bigg]
\bigg[1 + \sqrt{\frac{2}{5(1-3r^2/5R(t)^2)}}\bigg]\;.
\end{equation}
From this expression it can be observed that the radial pressure cancels out at the surface. On the other hand, notice that it is negative within the material distribution. 
The tangential pressure, from the $(\hat{2},\hat{2})$ (and $(\hat{3},\hat{3})$) components, takes the form
\begin{equation}
p_\perp(t,r) = - \frac{5}{12\pi(t_c-t)^2}
\bigg[1 - 2\Big( \frac{r}{R(t)} \Big)^{\!2} 
+ \frac{9r^4-23R(t)^2r^2+10R(t)^4}{5\sqrt{10}R(t)^4(1-3r^2/5R(t)^2)^{3/2}}\bigg]\;.
\end{equation}
Unlike radial pressure, the tangential component is negative in the innermost region of the distribution and becomes positive in the outermost region (see figure \ref{fig:Solution_Matter}).

Using the pressure components, the anisotropy factor is defined as $\Delta = p_\perp - p_r$. This quantity is very useful in characterising the material distribution and reveals more clearly the interplay between the various pressure components and their influence on the dynamics of the system. The pressure anisotropy factor is given by
\begin{align}
\Delta p &= p_\perp - p_r \nonumber \\
&= - \frac{5}{12\pi(t_c-t)^2}
\bigg[ \frac{2(r^2-R(t)^2)}{\sqrt{10}R(t)^2(1-3r^2/5R(t)^2)^{1/2}}
-\Big( \frac{r}{R(t)} \Big)^{\!2} \\
& \hspace{.5cm} +  \frac{9r^4-23R(t)^2r^2+10R(t)^4}{5\sqrt{10}R(t)^4(1-3r^2/5R(t)^2)^{3/2}} 
\bigg]\;.    \nonumber
\end{align}
Our calculations show that $\Delta > 0$ throughout the collapse, indicating that the tangential pressure exceeds the radial pressure. This anisotropy is a direct consequence of the spherical symmetry and the dynamic nature of the collapse.

\subsubsection{Exterior Region}

In the exterior region ($r > R(t)$), the matter content corresponds to Schwarzschild solution $T_{\mu\nu} = 0$ (vacuum). In this case, all energy conditions are trivially satisfied. Therefore we will only consider in detail the analysis of the energy conditions within the configuration.

\subsection{Energy Condition Analysis}
\label{subsec:energy_conditions}

We examine four fundamental energy conditions that constrain the behavior of matter in general relativity:

\begin{enumerate}
\item \textbf{Weak Energy Condition (WEC)}: $\rho \geq 0$ and $\rho + p_i \geq 0$ for any pressure $p_i$.
\item \textbf{Null Energy Condition (NEC)}: $\rho + p_i \geq 0$ for any pressure $p_i$. 
\item \textbf{Dominant Energy Condition (DEC)}: $\rho^2 \geq p^2_i$ for any pressure $p_i$.
\item \textbf{Strong Energy Condition (SEC)}: $\rho + p_r + 2p_\perp \geq 0$ and $\rho + p_i \geq 0$ for any pressure $p_i$.
\end{enumerate}
For anisotropic matter, these conditions must be evaluated separately for both the radial and tangential pressures. The most restrictive constraints typically arise from the radial pressure due to its negative values during collapse.

Before examining specific violations, we note that energy conditions in gravitational collapse must be interpreted carefully. These conditions were originally formulated for static matter configurations and their extension to highly dynamical systems involves conceptual subtleties. The violations we observe may indicate either unphysical matter or limitations in applying these classical constraints to collapse dynamics. We examine the four fundamental energy conditions that constrain the behavior of matter in general relativity:

\subsubsection{Weak Energy Condition}

The WEC requires both $\rho \geq 0$ and $\rho + p_i \geq 0$. While the energy density remains positive throughout the collapse, the condition $\rho + p_r \geq 0$ is violated due to the strongly negative radial pressure. 
On the other hand, it is observed that condition $\rho + p_\perp \geq 0$ is violated in the innermost region and that there is always a value $r_s$ where it begins to be satisfied. 
Our numerical analysis shows systematic violation of this condition across all collapse scenarios.

\subsubsection{Null Energy Condition}

From the previous paragraph we can see that the NEC, requiring $\rho + p_r \geq 0$ and $\rho + p_\perp \geq 0$, is also systematically violated. The violation occurs primarily through the radial pressure term, as $p_r$ becomes sufficiently negative to overcome the positive energy density contribution.

\subsubsection{Dominant Energy Condition}

The DEC requires $\rho^2 \geq p_r^2$ and $\rho^2 \geq p_\perp^2$. Given the large negative values of $p_r$ and the moderate positive values of $\rho$, this condition is violated throughout most of the collapse evolution. The violation becomes more pronounced as the collapse progresses and the pressure anisotropy increases.

\subsubsection{Strong Energy Condition}

The SEC requires $\rho + p_i \geq 0$ and $\rho + p_r + 2p_\perp \geq 0$. The latter constraint is most severely violated in the innermost region. However, it can also be seen to become positive earlier. This is caused by the effects of anisotropy on the system.

Our analysis demonstrates that all standard energy conditions are violated during the gravitational collapse process. The violations are characterized by a positive energy density $\rho > 0$, which is physically expected. The model also shows a strongly negative radial pressure $p_r < 0$ as a primary source of energy condition violations but a positive anisotropy factor $\Delta > 0$; this is a consequence of the geometry and non homogenous properties of matter the distribution.

\begin{figure*}[htbp]
\centering
\begin{tabular}{ccc}
\hspace{-1.1cm}
\includegraphics[width=0.36\textwidth]{Density.pdf} &
\hspace{-.6cm}
\includegraphics[width=0.36\textwidth]{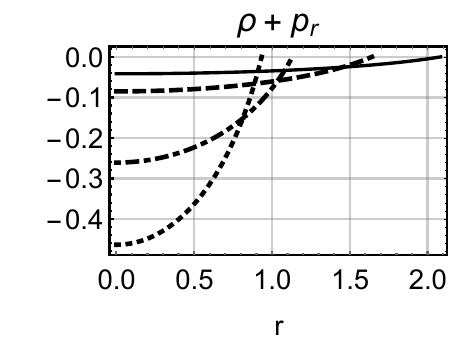} & \hspace{-.6cm}
\includegraphics[width=0.36\textwidth]{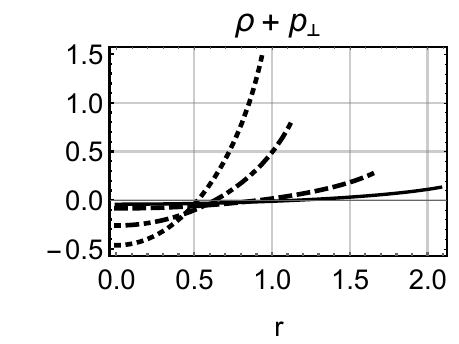} \\
\hspace{-1.1cm}
\includegraphics[width=0.36\textwidth]{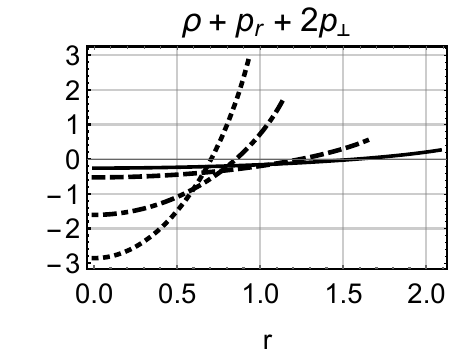} & \hspace{-.6cm}
\includegraphics[width=0.36\textwidth]{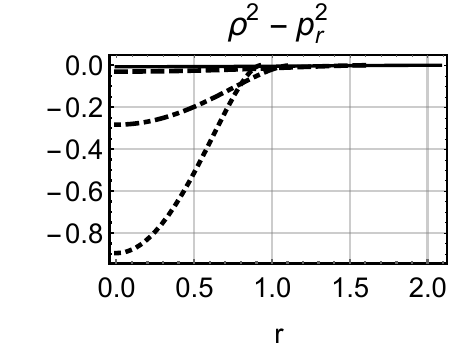} & \hspace{-.6cm}
\includegraphics[width=0.36\textwidth]{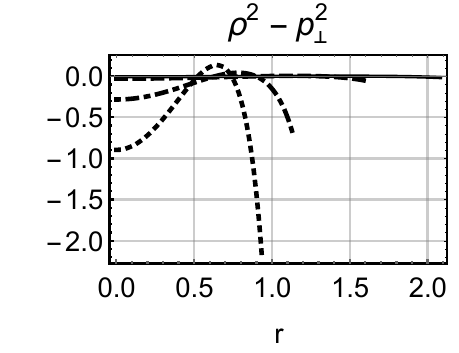}
\end{tabular}
\caption{Energy conditions. Aside from $\rho\geq 0$, all the constraints for the energy conditions are violated. Parameters used, $M_0=1$. The selected values are $t=0$ (solid line), $t=0.3 t_c$ (dashed line), $t=0.6t_c$ (dot--dashed line), $t=0.7t_c$ (short--dashed line).}
\label{fig:Energy-Conditions}
\end{figure*}

\subsection{Physical Interpretation and Validity}

The systematic violation of energy conditions in our model provides an opportunity to examine the physical foundations of gravitational collapse theory and the role of classical constraints in highly dynamical spacetimes. While these violations require careful interpretation, they do not invalidate the model's geometric insights or its contribution to understanding collapse dynamics.

\subsubsection{Energy Conditions in Dynamical Contexts}

Energy conditions were originally formulated as constraints on matter in equilibrium or slowly evolving configurations. Their extension to gravitational collapse involves several important considerations that help contextualize our results.

In highly dynamical systems, the relevant physics often involves effects that are not captured by pointwise constraints. The negative radial pressure in our model represents an instantaneous local property required to maintain the geometric structure of the collapse, while global quantities like total mass remain conserved throughout the evolution. This suggests that the violations may reflect the geometric demands of the specific collapse scenario rather than fundamental physical inconsistencies \cite{Hawking:1973uf,Wald:1984rg,Barcelo:2002bv}.

The extreme pressure anisotropy observed throughout the collapse emerges directly from Einstein's equations applied to our spherically symmetric geometry, rather than being imposed through artificial matter properties. This geometric origin indicates that the anisotropy represents a natural consequence of the gravitational dynamics we are modeling, providing insight into how spacetime curvature influences matter behavior during collapse.

The violations also highlight an important distinction between effective and fundamental matter properties. Our stress-energy tensor represents the effective matter content required to source the desired spacetime geometry, which may differ from the microphysical properties of realistic astrophysical matter. This approach parallels successful models in cosmology and modified gravity, where effective descriptions capture essential physics despite violating classical constraints.

\subsubsection{Model Contributions and Limitations}

Our model makes several valuable contributions to gravitational collapse theory while operating within clearly defined limitations. The unified Painlevé-Gullstrand framework provides exact analytical solutions for the complete collapse process, including detailed horizon dynamics and the transition between interior and exterior geometries. The identification of double apparent horizon phases adds to our understanding of the rich geometric structures possible during gravitational collapse, with precedents in other studies of anisotropic and inhomogeneous systems \cite{Ivanov:2002xf,Sharma:2012vc,Das:2020rfz}.

The model successfully demonstrates how geometric matching conditions constrain the collapse evolution and determine critical parameters for black hole formation. The relationship between initial compactness and horizon formation timescales provides physical intuition for the conditions necessary for successful collapse to a black hole.

However, the phenomenological nature of our energy density ansatz means the model should be interpreted as a geometric framework for understanding collapse dynamics rather than a direct description of specific astrophysical systems. The energy condition violations indicate that realistic collapse scenarios would require more sophisticated matter descriptions, possibly including quantum corrections or modified gravitational physics near the horizon.

The classical framework we employ becomes questionable during the final collapse phases, where quantum effects and the breakdown of classical spacetime concepts are expected. Nevertheless, the model provides valuable insights into the geometric aspects of horizon formation and the mathematical structure of collapse solutions.

\subsubsection{Physical Validity and Interpretation}

Despite the energy condition violations, several factors support the model's validity as a theoretical framework for understanding gravitational collapse. The mathematical self-consistency of the Einstein field equations, combined with smooth geometric matching and physically reasonable horizon formation, demonstrates that the model captures essential features of black hole formation.

The violations can be understood as arising from the constraints imposed by our idealized geometric setup rather than representing fundamental failures of general relativity. In realistic scenarios, departures from perfect spherical symmetry, the inclusion of rotation and magnetic fields, or the effects of quantum gravity could resolve these violations while preserving the essential collapse dynamics we have identified.

The model's predictions align with established results in several important respects: the final event horizon stabilizes at the Schwarzschild radius, the collapse occurs in finite proper time, and the geometric matching produces smooth spacetime evolution. These features provide confidence that the model captures the correct qualitative behavior despite its quantitative limitations.

Rather than viewing the energy condition violations as invalidating the model, they can be understood as indicating the boundaries of classical general relativity's applicability and pointing toward the physics needed for more complete collapse descriptions. The model thus serves both as a tool for understanding known collapse phenomena and as a guide for identifying where extensions of classical theory become necessary.

\section{Conclusions}
\label{sec:conclusions}
This work presents a specific analytical solution for gravitational collapse using Painlevé-Gullstrand coordinates, derived from a phenomenologically motivated energy density profile. The unified coordinate treatment eliminates technical difficulties while providing exact solutions for this particular collapse scenario.

Our analysis reveals several important features of anisotropic gravitational collapse. The model successfully reproduces expected black hole formation with the event horizon stabilizing at the Schwarzschild radius, while uncovering rich intermediate dynamics including double apparent horizon phases. These geometric structures, while perhaps surprising, have precedents in the literature and demonstrate the complexity possible in non-trivial collapse scenarios.

The critical parameter relationships we identify provide physical insight into the conditions necessary for black hole formation. The threshold initial compactness $R_0 = 25M_0/12$ represents a natural boundary in parameter space, separating configurations that form horizons immediately from those requiring finite evolution time. This relationship connects the initial geometric setup to the ultimate fate of the collapse, offering guidance for understanding realistic astrophysical scenarios.

The systematic energy condition violations we observe, rather than invalidating the model, illuminate the boundaries of classical general relativity's applicability to extreme gravitational systems. These violations arise from the geometric constraints of our idealized setup and point toward the need for either more sophisticated matter descriptions or recognition of where quantum corrections become essential. The model thus serves both as a tool for understanding classical collapse dynamics and as a guide for identifying where extensions of the theory become necessary.

The double apparent horizon structure and the geometric origin of pressure anisotropy represent genuine theoretical insights that advance our understanding of gravitational collapse beyond the classical Oppenheimer-Snyder framework. The exact analytical solutions we provide offer valuable benchmarks for numerical studies and serve as starting points for more realistic models incorporating rotation, magnetic fields, or quantum effects.

Future investigations could profitably explore whether the energy condition violations persist under more realistic matter descriptions, extend the analysis to include quantum corrections near the horizon, and investigate the stability of the horizon dynamics under perturbations. The framework developed here provides a solid foundation for such extensions while contributing to our understanding of black hole formation as a fundamental process in general relativity.

The balance between mathematical rigor and physical insight achieved in this model demonstrates the value of analytical approaches to gravitational collapse, even when idealized assumptions are necessary. By clearly delineating both the contributions and limitations of our approach, we provide a framework that advances theoretical understanding while pointing toward the physics needed for more complete descriptions of these fascinating phenomena.

\section*{Acknowledgments}

\bibliography{biblio}
\bibliographystyle{unsrt}

\end{document}